\documentclass[lettersize,journal]{IEEEtran}
\usepackage{amsmath,amsfonts}
\usepackage{algorithmic}
\usepackage{array}
\usepackage[caption=false,font=normalsize,labelfont=sf,textfont=sf]{subfig}
\usepackage{textcomp}
\usepackage{stfloats}
\usepackage{url}
\usepackage{verbatim}
\usepackage{graphicx}
\usepackage{mathrsfs}
\usepackage{algorithm}
\usepackage{amssymb}
\usepackage{cite}
\usepackage{epstopdf}
\usepackage{epsfig}
\usepackage{xcolor}
\usepackage{booktabs}
\usepackage{bm}

\newtheorem{definition}{Definition}

\newtheorem{remark}{Remark}
\newtheorem{theorem}{Theorem}
\newtheorem{lemma}{Lemma}
\newtheorem{assumption}{Assumption}
\newtheorem{corollary}{Corollary}

\def\BibTeX{{\rm B\kern-.05em{\sc i\kern-.025em b}\kern-.08em
    T\kern-.1667em\lower.7ex\hbox{E}\kern-.125emX}}
\usepackage{balance}
\begin{document}
\title{Event-triggered privacy preserving consensus control with edge-based additive noise}
\author{Limei Liang, Ruiqi Ding, Shuai Liu
\thanks{Manuscript created March, 2023. \emph{(Corresponding author: Shuai Liu.)}

Limei Liang and Shuai Liu are with the School of Control Science and Engineering, Shandong University, Jinan 250061, China (E-mail: lianglimei@mail.sdu.edu.cn; liushuai@sdu.edu.cn).

Ruiqi Ding is with the School of Mathematics, Shandong University, Jinan 250100, China (E-mail:202022171217@mail.sdu.edu.cn).}}

\maketitle

\begin{abstract}
	In this article, we investigate the distributed privacy preserving weighted consensus control problem for linear continuous-time multi-agent systems under the event-triggering communication mode. A novel event-triggered privacy preserving consensus scheme is proposed, which can be divided into three phases. First, for each agent, an event-triggered mechanism is designed to determine whether the current state is transmitted to the corresponding neighbor agents, which avoids the frequent real-time communication. Then, to protect the privacy of initial states from disclosure, the edge-based mutually independent standard white noise is added to each communication channel. Further, to attenuate the effect of noise on consensus control, we propose a stochastic approximation type protocol for each agent. By using the tools of stochastic analysis and graph theory, the asymptotic property and convergence accuracy of consensus error is analyzed. Finally, a numerical simulation is given to illustrate the effectiveness of the proposed scheme.
\end{abstract}


\section{Introduction}
In recent years, the study of distributed consensus control problem of multi-agent systems (MASs) have received increasing attention, due to its broad applications in multi-robot coordination \cite{Tas22}, autonomous underwater vehicles \cite{Xu20}, smart grid \cite{Zhao18}, unmanned aerial vehicle formation \cite{Dong15} and other fields \cite{Qin17,Ma17,Ding18}. In general, traditional consensus algorithms require each agent to share its local state, or even its local input, with neighboring agents. This means that an agent's exact information could be monitored by some malicious adversaries, and such privacy disclosure is very dangerous. Therefore, from the perspective of network security, the designed consensus algorithm should be secure and able to prevent the privacy of each agent from being leaked, which brings a new challenge to the consensus control research of MASs. To achieve the consensus while protecting the privacy of agents from disclosure, the researchers have conducted extensive research on the privacy preserving consensus control problem. Consequently, considerable progress has been made in this direction.

Generally speaking, the related research on privacy protection can be roughly divided into four categories, namely, cryptology-based approach, state decomposition-based approach, output mask-based approach and random noise or deterministic perturbation signal-based approach. To mention a few, in \cite{Ruan19}, the authors proposed a privacy preserving method based on homomorphic encryption for undirected networks, which can guarantee the security and privacy of an agent as long as it has at least one legitimate neighbor. Further, in \cite{Had20}, a ratio consensus algorithm combined with homomorphic encryption is proposed. In \cite{Wang19}, the author proposed a state decomposition based privacy-preserving average consensus method. Furthermore, in \cite{Zhang22}, this method is extended to the privacy preserving dynamic average consensus problem. In \cite{Hu22}, the authors proposed a privacy protection scheme based on the output masking. In \cite{Mo17}, a random noise-based privacy preserving average consensus scheme is proposed. In \cite{Xiong22}, a deterministic perturbation signals-based privacy preserving average consensus scheme is proposed.

In all the above privacy-preserving consensus algorithms, the communication between agents occurs frequently in each iterative step. Considering this time-triggered communication mode will lead to the improvement of communication network utilization and a large loss of computing resources, relevant scholars have introduced the event-triggered communication mechanism into the design of privacy protection scheme. To list a few, in \cite{Gao19}, an event-triggered differential privacy scheme is proposed to protect the initial states of MASs from disclosure. In \cite{Wang21}, a kind of event-triggered privacy-preserving consensus algorithm based on output mask is proposed for MASs. In \cite{Yang22}, an event-triggered privacy-preserving bipartite consensus algorithm based on encryption is proposed for discrete-time nonlinear MASs.

Since the privacy protection scheme based on adding random noise is simple to design and easy to implement, most of the existing privacy preserving schemes make privacy protection by adding random noise to the transmitted information. They do this using a tool called differential privacy. Relevant results can be referred to \cite{Nozari17,Fiore19,Wang19,He20,Dong20}. In these schemes, each agent adds a random noise to its original state, and then transmits this obfuscated signal to all its neighbors. Since each agent usually has a different number of neighbors, the above scheme does not fully exploit the heterogeneity of MASs, which will damage the privacy protection performance. In addition, these works only focus on discrete-time MASs, while there are few related research results on continuous-time ones.

Inspired by the above works, in this paper, we aim to propose an event-triggered privacy preserving consensus control scheme with edge-based additive noise for continuous-time MASs. Specifically, we first design an event-triggered communication mechanism to determine whether the current state is transmitted to the corresponding neighbor agents, which avoids the frequent real-time communication. Then, we develop a novel privacy preserving scheme, which ensures the states of agents are not disclosed by adding edge-based mutually independent standard white noises to the communication channels with neighbors. Furthermore, we propose a stochastic approximation type protocol for each agent to attenuate the effect of noise on consensus as much as possible. 

The main contribution of this paper is reflected in the following three aspects. 

1) An appropriate event triggering mechanism is proposed, which avoids frequent communication and prevents Zeno behavior.

2) A privacy preserving method of adding edge-based random noise is presented, which improves the privacy preserving performance and brings more degrees of design freedom.

3) The proposed privacy preserving consensus control scheme is truly distributed, which is convenient to implement for MASs.

The remainder of this article is structured as follows. Section \ref{Pro} introduces the preliminaries of graph theory and makes the problem statement. Section \ref{Main} gives the main results of on consensus control and analyzes the convergence accuracy and Zeno behavior. Section \ref{Sim} presents a numerical simulation to verify the effectiveness and section \ref{Con} concludes this paper.

{\bf Notations:} 
$\mathbb{R}$, $\mathbb{R}_{\ge 0}$ and $\mathbb{C}$ denote the set of real numbers, non-negative real numbers and complex numbers, respectively. $\mathbb{C}^{n}$ denotes the $n$-dimensional complex vectors. $\mathbb{R}^{m\times n}$ and $\mathbb{C}^{m\times n}$ denote the set of $m\times n$ real and complex matrices, respectively. 
The superscript $T$ denotes the transpose of a matrix or vector, and the superscript $-1$ denotes the inverse of a matrix. $\text{diag}(\cdot)$ denotes a diagonal matrix. $|\cdot|$ denotes the absolute value of a scalar. Let $\mathbf{1}_{N}$ and $\mathbf{0}_{N}$ be $n$-dimensional column vectors with all elements respectively being $1$ and $0$.  For a vector $x$, $\|x\|$ denotes the Euclidean norm. For a random variable $X$, $\mathbb{E}(X)$ and $\mathbb{V}(X)$ indicate the expectation and variance, respectively. The notion $f(t)=O(g(t))$ denotes $\text{lim sup}_{t\to\infty}|f(t)/g(t)|<\infty$ and $f(t)=o(g(t))$ denotes $\text{lim sup}_{t\to\infty}|f(t)/g(t)|=0$.

\section{Problem Statement}\label{Pro}
In this paper, we consider the privacy preserving consensus control problem for a MAS consisting of $N$ interconnected agents. The communication topology among agents can be represented by a digraph $\mathcal{G}$, which consists of a node set $\mathcal{V}=\{1,2,\cdots,N\}$ and an edge set $\mathcal{E}=\{(j,i):i,j\in \mathcal{V}\}$, where $(j,i)$ means that node $i$ can receive information from node $j$, but not necessarily vice versa. For this digraph $\mathcal{G}$, the adjacency matrix $\mathcal{A}=[a_{ij}]\in \mathbb{R}^{N\times N}$ is defined as $a_{ii}=0$, $a_{ij}>0$ if $(j,i)\in \mathcal{E}$ and $a_{ij}=0$ otherwise, and the Laplacian matrix $\mathcal{L}=[l_{ij}]\in \mathbb{R}^{N\times N}$ is defined as $l_{ii}=\sum_{j\ne i}a_{ij}$, $l_{ij}=-a_{ij}$ for $i\ne j$. For each $i\in\cal{V}$, we define the set of in-neighbors of node $i$ as $\mathcal{N}_{i}=\{j \in \mathcal{V}:(j,i)\in\mathcal{E}, i \neq j \}$. 

The dynamics of the $i$-th agent is described by
\begin{equation}\label{system1} 
	\dot{x}_{i}(t)=u_{i}(t),
\end{equation}
where $x_{i}(t)\in\mathbb{R}$ is the internal state of agent $i$ and $u_{i}(t)\in\mathbb{R}$ is the control input to be designed.

The objective of this paper is to design a distributed event-triggered control law for each agent to protect the agents' states from disclosure while the individual agents are able to collaborate with each other to achieve consensus. In order to derive our main results, the following assumption and lemma are necessary.
\begin{assumption} \label{assumption1}
	\rm The digraph $\mathcal{G}$ contains a spanning tree.
\end{assumption}

\begin{lemma}\cite{Yang16}\label{lemma1}
	If digraph $\mathcal{G}$ contains a spanning tree, $0$ is a simple eigenvalue of $\mathcal{L}$, with $\mathbf{1}_{N}$ is the right eigenvector and $r^{T}=[r_{1},\cdots,r_{N}]$ is the left eigenvector, satisfying $r^{T}\mathbf{1}_{N}=1$. Further, all the other eigenvalues of $\mathcal{L}$ have positive real parts.
\end{lemma}

\section{Main Results}\label{Main}
In this section, we introduce a privacy preserving event-triggered control scheme for each agent to solve the consensus problem and prove that the Zeno behavior can be excluded.
\subsection{Controller design}
It is assumed that each agent can monitor its own state continuously and decide when to transmit its current state over the network based on the event-triggered mechanism.

For each agent $i$, define $\tilde{x}_{i}(t)=x_{i}(t_{k,i}), \forall t \in [t_{k,i},t_{k+1,i}]$ to denote the latest transmitted state of agent $i$, where $t_{k,i}$ denotes the $k$-th event-triggering instant of agent $i$, which will be determined later by the designed event-triggered mechanism.

To protect the privacy of initial states from disclosure, the edge-based mutually independent standard white noise is added to each communication channel. The information of the $j$-th agent received by the $i$-th agent is
\begin{equation*}\label{y_ji} 
	\tilde{y}_{ji}(t)=\tilde{x}_{j}(t)+\sigma_{ji}\eta_{ji}(t),
\end{equation*}
where $\eta_{ji}(t), i\in \mathcal{V}, j\in\mathcal{N}_{i}$ are mutually independent standard white noises, and $\sigma_{ji}>0$ is the noise intensity. Here, the noises are independent of the initial states.

Based on the information from the neighbors, the control law of agent $i$ is given as
\begin{equation}\label{ui}
	u_{i}(t)=a(t)\sum_{j\in\mathcal{N}_{i}}a_{ij}(\tilde{y}_{ji}(t)-\tilde{x}_{i}(t)),
\end{equation}
where the consensus gain function $a(t):[0,\infty)\to [0,\infty)$ is a decreasing and bounded piecewise continuous function, which satisfies the following assumption.
\begin{assumption} \label{assumption2}
	\rm $a(t)$ satisfies that $\lim_{t \to \infty} t^{\gamma}a(t)$ exists and is positive, where $\gamma \in (0.5,1)$.
\end{assumption}
\begin{remark}
	Assumption \ref{assumption2} guarantees that $\int_{0}^{\infty}a(s)ds=\infty$ and $\int_{0}^{\infty}a^{2}(s)ds<\infty$, which not only avoids the failure of controller caused by too fast attenuation rate, but also can suppress the noise to make the system converge  eventually.
\end{remark}
\begin{remark}
	A direct consequence of Assumption 2 is that $\forall t>0$, $\exists~\underline{C},~\overline{C}>0$ makes
	\begin{equation*}
		\frac{\underline{C}}{(1+t)^{\gamma}} \le a(t) \le \frac{\overline{C}}{(1+t)^{\gamma}}.
	\end{equation*}
\end{remark}

We present the design of the event-triggered mechanism here. For each agent $i$, denote the state error as
\begin{equation}\label{error}
	e_{i}(t)=\tilde{x}_{i}(t)-x_{i}(t).
\end{equation}

Then, the event-triggered mechanism for agent $i$ is designed as
\begin{equation}\label{event1}
	t_{k+1,i}=\inf\{t>t_{k,i}:e^{T}_{i}(t)e_{i}(t)\ge a(t)\}.
\end{equation}

Substituting \eqref{ui} into \eqref{system1}, the dynamics of agent $i$ in the form of $\rm It\hat{o}$ stochastic differential equation is given by
\begin{equation}\label{xi}
	\begin{aligned}
		dx_{i}(t)
		&=a(t)\sum_{j\in\mathcal{N}_{i}}a_{ij}(\tilde{y}_{ji}(t)-\tilde{x}_{i}(t))dt\\
		&=a(t)\sum_{j\in\mathcal{N}_{i}}a_{ij}(\tilde{x}_{j}(t)-\tilde{x}_{i}(t))dt\\
		&\quad+a(t)\sum_{j\in\mathcal{N}_{i}}a_{ij}\sigma_{ji}\eta_{ji}(t)dt\\
		&=-a(t)\sum_{j=1}^{N}l_{ij}(x_{j}(t)+e_{j}(t))dt\\
		&\quad+a(t)\sum_{j\in\mathcal{N}_{i}}a_{ij}\sigma_{ji}dw_{ji}(t),
	\end{aligned}
\end{equation}
where $w_{ji}(t)$ is a one dimensional standard Brownian motion.

Let $x(t)=[x_{1}(t),\cdots,x_{N}(t)]^{T}$ and $e(t)=[e_{1}(t),\cdots,e_{N}(t)]^{T}$, the dynamics of MAS can be described as
\begin{equation}\label{x}
	dx(t)=-a(t)\mathcal{L}(x(t)+e(t))dt+a(t)\varSigma dw(t).
\end{equation}
Here, $\varSigma=$\rm{diag}$(\alpha_{1}\varSigma_{1},\cdots,\alpha_{N}\varSigma_{N})$, where $\alpha_{i}$ is the $i$-th row of the weighted adjacency matrix $\mathcal{A}$ and $\varSigma_{i}=$\rm{diag}$(\sigma_{1i},\cdots,\sigma_{Ni})$ with $\sigma_{ji}=0$ for $j\notin\mathcal{N}_{i}$. $w(t)=[w_{11}(t),\cdots,w_{N1}(t),\cdots,w_{NN}(t)]^{T}$ is an $N^{2}$ dimensional standard Brownian motion. Let $\mathcal{F}_{t}=\sigma\left\{x(s):s\le t\right\}$ be the filtration generated by $x(t)$.

Next, we will focus on the consensus problem. To this end, denote $J=\mathbf{1}_{N}r^{T}$ and define the disagreement vector as
\begin{equation}\label{delta}
	\delta(t)=(I_{N}-J)x(t).
\end{equation}
Since $(I_{N}-J)\mathcal{L}=\mathcal{L}=\mathcal{L}(I_{N}-J)$, the dynamics of $\delta(t)$ is
\begin{equation*}\label{ddelta}
	d\delta(t)=-a(t)\mathcal{L}(\delta(t)+e(t))dt+a(t)(I_{N}-J)\varSigma dw(t).
\end{equation*}

According to Lemma \ref{lemma1}, there are matrices $T=[\mathbf{1}_{N}\  Y]$ and $T^{-1}=[r\ W^{T}]^{T}$, where $Y\in\mathbb{C}^{N\times(N-1)}$ and $W\in\mathbb{C}^{(N-1)\times N}$, such that
\begin{equation}\label{Jordan}
	T^{-1}\mathcal{L}T=J_{\mathcal{L}}=\begin{bmatrix}
		0 & \mathbf{0}_{N-1}^{T}\\
		\mathbf{0}_{N-1} & \tilde{\mathcal{L}}
	\end{bmatrix},
\end{equation}
where $J_{\mathcal{L}}$ is the Jordan canonical form of the matrix $\mathcal{L}$ and $\tilde{\mathcal{L}}$ is the block diagonal matrix in which the diagonal entries are the nonzero eigenvalues of $\mathcal{L}$. 

Using the state transformation, we introduce a new variable
\begin{equation*}\label{varepsilon}
	\varepsilon(t)=T^{-1}\delta(t)=[\varepsilon_{1}(t), \tilde{\varepsilon}^{T}(t)]^{T},
\end{equation*}
where $\varepsilon_{1}(t)\in\mathbb{C}$ and $\tilde{\varepsilon}(t)=[\varepsilon_{2}(t),\cdots,\varepsilon_{N}(t)]^{T}\in\mathbb{C}^{N-1}$.

Since $r^{T}\mathbf{1}_{N}=1$, it can be seen from \eqref{delta} that
\begin{equation*}\label{varepsilon1}
	\varepsilon_{1}(t)=r^{T}\delta(t)\equiv 0,
\end{equation*}
and $\tilde{\varepsilon}(t)$ satisfies
\begin{equation}\label{dvarepsilon2}
	\begin{aligned}		
		d\tilde{\varepsilon}(t)=&-a(t)\tilde{\mathcal{L}}\tilde{\varepsilon}(t)dt-a(t)\tilde{\mathcal{L}}We(t)dt\\
		&+a(t)W(I_{N}-J)\varSigma dw(t).
	\end{aligned}
\end{equation}

Then, the consensus problem can be converted to the convergence problem of $\delta(t)$, which is equivalent to the convergence problem of $\tilde{\varepsilon}(t)$. For further analysis, we present here the following auxiliary lemma regarding $a(t)$, the proof of which can be found in the Appendix.
\begin{lemma}\label{a1}
	It is supposed that Assumption \ref{assumption2} holds. Then, for any given $\mu>0, p>2$, we have
	\begin{equation}\label{lemma2_1}
		e^{-\mu\int_0^ta(s)ds}=o(t^{-\frac{p\gamma}{2}}).
	\end{equation}
	and
	\begin{equation}
		\lim_{t\to\infty}t^{\frac{p\gamma}{2}}\int_0^t a^{\frac{p+2}{2}}(s)e^{-\mu\int_s^ta(r)dr}ds=\frac{(\lim_{t\to\infty}t^{\gamma}a(t))^{\frac{p}{2}}}{\mu}.
	\end{equation}
\end{lemma}

\subsection{Convergence analysis}
Next, we will conduct the convergence analysis of $\tilde{\varepsilon}(t)$.
The relevant result is given by the following theorem.

\begin{theorem}\label{theorem1}
	For a MAS with communication topology $\mathcal{G}$, it is supposed that Assumption \ref{assumption1} holds. Apply the protocol \eqref{ui} to the system \eqref{system1}. If Assumption \ref{assumption2} holds, then $\tilde{\varepsilon}(t)$ converges in the $p$-th moment, which means that $\lim_{t\to\infty}\mathbb{E}(x_j(t)-x_i(t))^p=0$ for $\forall p\ge 2$.
\end{theorem}
{\bf Proof}.
Since for $p>2$, the convergence of $\tilde{\varepsilon}(t)$ in the $p$-th moment implies the convergence in mean square, we can assume $p>2$ in the following proof.

Choose a Lyapunov function $V(t)=(\tilde{\varepsilon}^T(t)\tilde{\varepsilon}(t))^{\frac{p}{2}}$. It follows from \eqref{dvarepsilon2} that
\begin{equation*}
	\begin{aligned}
		dV(t)&=p(\tilde{\varepsilon}^T(t)\tilde{\varepsilon}(t))^{\frac{p-2}{2}}\tilde{\varepsilon}^T(t)d\tilde{\varepsilon}(t)\\
		&\quad+\frac{p}{2}a^2(t)(\tilde{\varepsilon}^T(t)\tilde{\varepsilon}(t))^{\frac{p-2}{2}}\\
		&\quad\times\text{trace}(\varSigma^T(I_{N}-J)^TW^TW(I_{N}-J)\varSigma)dt\\
		&\quad +\frac{p(p-2)}{2}a^2(t)(\tilde{\varepsilon}^T(t)\tilde{\varepsilon}(t))^{\frac{p-4}{2}}\text{trace}(\tilde{\varepsilon}(t)\tilde{\varepsilon}^T(t)\\
		&\quad W(I_{N}-J)\varSigma\varSigma^T(I_{N}-J)^TW^T)dt\\
		&=-pa(t)V^{\frac{p-2}{p}}(t)\tilde{\varepsilon}^T(t)\tilde{\mathcal{L}}\tilde{\varepsilon}(t)dt\\
		&\quad-pa(t)V^{\frac{p-2}{p}}(t)\tilde{\varepsilon}^T(t)\tilde{\mathcal{L}}We(t)dt\\
		&\quad+pa(t)V^{\frac{p-2}{p}}(t)\tilde{\varepsilon}^T(t)W(I_{N}-J)\varSigma dw(t)\\
		&\quad +\frac{p}{2}a^2(t)V^{\frac{p-2}{p}}(t)\left \| W(I_{N}-J)\varSigma\right \|_{F}^2 dt\\
		&\quad+\frac{p(p-2)}{2}a^2(t)V^{\frac{p-4}{p}}(t)\\
		&\quad \tilde{\varepsilon}^T(t)W(I_{N}-J)\varSigma\varSigma^T(I_{N}-J)^TW^T\tilde{\varepsilon}(t)dt\\
		& \le -p\tilde{\lambda}_{\min}a(t)V(t)dt\\
		&\quad-pa(t)V^{\frac{p-2}{p}}(t)\tilde{\varepsilon}^T(t)\tilde{\mathcal{L}}We(t)dt\\
		&\quad+\bigg(\frac{p}{2}\left \| W(I_{N}-J)\varSigma\right \|_{F}^2 +\frac{p(p-2)}{2}\tilde{\mu}_{max}\bigg)\\
		&\quad\times a^2(t)V^{\frac{p-2}{p}}(t)dt\\
		& \quad +pa(t)V^{\frac{p-2}{p}}(t)\tilde{\varepsilon}^T(t)W(I_{N}-J)\varSigma dw(t),
	\end{aligned}
\end{equation*}
where $\tilde{\lambda}_{\min}$ is the minimal eigenvalue of $\frac{\tilde{\mathcal{L}}+\tilde{\mathcal{L}}^T}{2}$, $\tilde{\mu}_{\max}$ is the maximal eigenvalue of $W(I_{N}-J)\varSigma\varSigma^T(I_{N}-J)^TW^T$.

Choose two constants $\theta_{1}, \theta_{2}$ satisfying
\begin{equation*}
	\begin{aligned}
		C_{0}\overset\Delta=& p\tilde{\lambda}_{\min}-(p-1)\theta_{1}^{\frac{p}{p-1}}-\frac{p-2}{2}\theta_{2}^{\frac{p}{p-2}}\\
		&\qquad\qquad\ \big(\left \| W(I_{N}-J)\varSigma\right \|_{F}^2+(p-2)\tilde{\mu}_{max}\big)>0.
	\end{aligned}
\end{equation*}
Then by Young's inequality and the definition of operator norm, we have  
\begin{equation*}
	\begin{aligned}	
		&-pa(t)V^{\frac{p-2}{p}}(t)\tilde{\varepsilon}^T(t)\tilde{\mathcal{L}}We(t)\\
		\le &pa(t)\bigg(  \frac{p-1}{p} \theta_{1}^{\frac{p}{p-1}}V(t)+\frac{1}{p}\theta_{1}^{-p}\|\tilde{\mathcal{L}}W\|^p\|e(t)\|^p \bigg),
	\end{aligned}	
\end{equation*}
\begin{equation*}
	a^2(t)V^{\frac{p-2}{p}} (t)\le \frac{p-2}{p}\theta_{2}^{\frac{p}{p-2}}a(t)V(t)+\frac{2}{p}\theta_{2}^{-\frac{p}{2}}a^{\frac{p+2}{2}}(t).
\end{equation*}
Therefore,
\begin{equation}
	\begin{aligned}	
		dV(t) \le &-C_{0}a(t)V(t)dt+C_{1}a^{\frac{p+2}{2}}(t)dt\\
		& +pa(t)V^{\frac{p-2}{p}}(t)\tilde{\varepsilon}^T(t)W(I_{N}-J)\varSigma dw(t),
	\end{aligned}
\end{equation}
\begin{equation}
	\mathbb{E}V(t) \le -C_{0}\int_{0}^{t}a(s)EV(s)ds+C_{1}\int_{0}^{t}a^{\frac{p+2}{2}}(s)ds,
\end{equation}
where $C_{1}=\theta_{1}^{-p}\|\tilde{\mathcal{L}}W\|^p\|e(t)\|^p+\big(\left \| W(I_{N}-J)\varSigma\right \|_{F}^2+(p-2)\tilde{\mu}_{max} \big) \theta_{2}^{-\frac{p}{2}}$.

By the comparision theorem, we have
\begin{equation}
	\begin{aligned}	
		\mathbb{E}V(t)&\le e^{-C_{0}\int_0^t a(s)ds}V(0)\\
		&\quad+C_1\int_0^t a^{\frac{p+2}{2}}(s)e^{-C_{0}\int_s^ta(r)dr}ds.
	\end{aligned}
\end{equation}
Further, according to Lemma \ref{a1}, we can get that
\begin{equation}\label{rate1}
	\mathbb{E}V(t) \le C_{2,p} t^{-\frac{p\gamma}{2}},
\end{equation}
where $C_{2,p}$ is a constant. Therefore,
\begin{equation*}
	\lim_{t \to \infty} \mathbb{E}V(t)=0,
\end{equation*}
which completes the proof.  $\hfill\blacksquare$

\subsection{Accuracy analysis}
For common weighted consensus algorithms, the state of each agent can converge to the weighted average of the initial states with certainty. However, due to the inherent property of the privacy protection mechanism, the proposed control algorithm cannot reach the weighted average of the initial states deterministically. Therefore, we will further analyze the statistical properties of the convergence value corresponding to the proposed algorithm.
\begin{definition}
	For any given initial state $x(0)$, $\rho\in(0,1)$, $\alpha\in\mathbb{R}_{\ge 0}$, a stochastic system is said to achieve $(\rho,\alpha)$ accuracy if the state of every agent converges in the mean-square sense to a random variable $x^*$ such that $\mathbb{P}\{|x^*-\mathbb{E}(x^*)|<\alpha\}\ge 1-\rho$.
\end{definition}
\begin{corollary}
	Under the proposed privacy preserving consensus scheme, each agent can achieve
	\begin{equation*}
		\Bigg(\rho,\sqrt{\frac{r^{T}\varSigma\varSigma^{T}r}{\rho(2\gamma-1)}}\Bigg)
	\end{equation*}
	accuracy and the mean of the convergence value $x_{\infty}$ is an unbiased estimate of their initial states' weighted average.
\end{corollary}
{\bf Proof}.
Multiply $J$ by both sides of \eqref{x}. Since $J\mathcal{L}=0$, we can get that
\begin{equation*}
	Jdx(t)=a(t)J\varSigma dw(t),
\end{equation*}
which means that
\begin{equation*}
	Jx(t)=Jx(0)+\int_{0}^{t}a(s)J\varSigma dw(s).
\end{equation*}
Thus, we can get that the covergence value $x_{\infty}=\lim_{t\to\infty}r^{T}x(t)$.

Since $w(t)=[w_{11}(t),\cdots,w_{N1}(t),\cdots,w_{NN}(t)]^{T}$ is an $N^{2}$ dimensional standard Brownian motion, we can further obtain that
\begin{equation*}
	\begin{aligned}
		\mathbb{E}(x_{\infty})&=r^Tx(0),\\
		\mathbb{V}(x_{\infty})&=\frac{r^{T}\varSigma\varSigma^{T}r}{2\gamma-1}.
	\end{aligned}
\end{equation*}
According to Chebyshev's inequality,
\begin{equation*}
	\mathbb{P}\{|x_{\infty}-\mathbb{E}(x_{\infty})|<\alpha\}\ge 1-\frac{\mathbb{V}(x_{\infty})}{\alpha^2}.
\end{equation*}
Choosing $\alpha=\sqrt{\frac{r^{T}\varSigma\varSigma^{T}r}{\rho(2\gamma-1)}}$, we can get that $\mathbb{P}\{|x_{\infty}-\mathbb{E}(x_{\infty})|<\alpha\}\ge 1-\rho$. This completes the proof.  $\hfill\blacksquare$

\subsection{Zeno behavior analysis}
The following result excludes the Zeno behavior a.s..
\begin{theorem}\label{theorem2}
	Consider system \eqref{system1} with the control law given in \eqref{ui}. Under the same conditions as in Theorem \ref{theorem1}, for any agent $i$, the following equality holds:
	\begin{equation}
		\mathbb{P}\{\inf_{k}( t_{k+1,i}-t_{k,i})=0\}=0,
	\end{equation}
	which means the lower bound for inter-execution time of agent $i$ is strictly positive a.s.. Therefore, the Zeno behavior can be excluded a.s..
\end{theorem}
{\bf Proof}.
Without loss of generality, we assume $a(t)=(t+1)^{-\gamma}$ in this proof.

Fix a constant $c >0$, for any agent $i$, define 
\begin{align*}
	\triangle t_{k,i}:&=t_{k+1,i}-t_{k,i}, \\
	d_{n,i}:&=\min_{0\le k\le n-1} \triangle t_{k,i},\\
	A_{n,i}:&= \left \{ t_{n,i}< \infty,\ \triangle t_{n,i} \le d_{n,i} \le c\right \}.
\end{align*}		
Then
\begin{equation}
	\begin{aligned}
		&\mathbb{P}\{\inf_{k}( t_{k+1,i}-t_{k,i})=0\}\\
		=&\mathbb{P}\{\exists\ \text{a decreasing subsequence}\ \triangle t_{k_{l},i}\ \text{converges to 0} \}\\
		\le&\mathbb{P}\{A_{n,i}\ \text{i.o.}\}.
	\end{aligned}
\end{equation}
\par To prove $\mathbb{P}\{\inf_{k}( t_{k+1,i}-t_{k,i})=0\}=0$, it suffices to show $\mathbb{P}\{A_{n,i}\ \text{i.o.}\}=0$. Further, according to Borel-Cantelli lemma, to prove $\mathbb{P}\{A_{n,i}\ \text{i.o.}\}=0$, it suffices to show
\begin{equation} \label{17}
	\sum_{n=1}^{\infty } \mathbb{P}\{A_{n,i}\} < \infty.
\end{equation}
Therefore, the remainder of this proof  aims to show (\ref{17}).
\par According to \eqref{error} and \eqref{xi}, for $t \in [t_{k,i},t_{k+1,i})$,
\begin{equation}
	\begin{aligned}
		d e_{i}(t) &=-d x_{i}(t) \\
		&=-a(t)\sum_{j\in\mathcal{N}_{i}}a_{ij}(\tilde{x}_{j}(t)-\tilde{x}_{i}(t))dt\\
		&\quad-a(t) \sum_{j \in \mathcal{N}_{i}} a_{i j} \sigma_{j i} dw_{ji}(t). \\
	\end{aligned}
\end{equation}
Then, we can get
\begin{equation}
	\begin{aligned}
		e_{i}(t)=&-\sum_{j \in \mathcal{N}_{i}} a_{i j}\int _{t_{k,i}}^{t}(\tilde{x}_{j}(t)-\tilde{x}_{i}(t))a(t)dt\\
		&-\sum_{j \in \mathcal{N}_{i}} a_{i j} \sigma_{j i} \int _{t_{k,i}}^{t}a(t)dw_{ji}(t),
	\end{aligned}
\end{equation}
and
\begin{equation}\label{trigger}
	\begin{aligned}
		e^{T}_{i}(t)e_{i}(t)
		\le &2\Bigg[N\sum_{j \in \mathcal{N}_{i}} a_{i j}^2 \bigg(\int _{t_{k,i}}^{t}\left\|\tilde{x}_{j}(t)-\tilde{x}_{i}(t)\right \|a(t)dt\bigg)^2\\
		&+\bigg(\sum_{j \in \mathcal{N}_{i}} a_{i j} \sigma_{j i}\int _{t_{k,i}}^{t}a(t)dw_{ji}(t)\bigg)^2\Bigg].
	\end{aligned}
\end{equation}

And then we derive the expression for $\mathbb{P}\{A_{n,i}\}$. For the sake of further derivation, we first define $\Pi_{i_1}(t)=2N\sum_{j \in \mathcal{N}_{i}} a_{i j}^2 \Big(\int _{t_{n,i}}^{t}\left\|\tilde{x}_{j}(t)-\tilde{x}_{i}(t)\right \|a(t)dt\Big)^2$ and $\Pi_{i_2}(t)=2 \Big(\sum_{j \in \mathcal{N}_{i}} a_{ij} \sigma_{j i}\int _{t_{n,i}}^{t}a(t)dw_{ji}(t)\Big)^2$. Then,
\begin{equation}
	\begin{aligned}
		&\quad \mathbb{P}\{A_{n,i}\}\\ &=\mathbb{P}\{( t_{n,i}< \infty,\ \triangle t_{n,i} \le d_{n,i} \le c)\}\\		
		& \le \mathbb{P}\{t_{n+1,i} \le t_{n,i}+d_{n,i}< \infty,\ d_{n,i}\le c\}\\ 
		&= \mathbb{P}\bigg\{\sup _{t_{n,i} \le t \le t_{n,i}+d_{n,i}}\frac{e^{T}_{i}(t)e_{i}(t)}{a(t)} \ge 1,\\
		&\qquad \qquad \qquad \qquad \qquad \qquad t_{n,i}< \infty,\ d_{n,i}\le c\bigg\}\\
		&\le  \mathbb{P}\bigg\{\sup _{t_{n,i} \le t \le t_{n,i}+d_{n,i}}\frac{\Pi_{i_1}(t)+\Pi_{i_2}(t)}{a(t)} \ge 1,\\
		&\qquad \qquad \qquad \qquad \qquad \qquad t_{n,i}<\infty,\ d_{n,i}\le c\bigg\} \\
		& \le  \mathbb{P}\bigg\{\sup _{t_{n,i} \le t \le t_{n,i}+d_{n,i}}\frac{\Pi_{i_1}(t)}{a(t)} \ge 1,\ t_{n,i}<\infty,\ d_{n,i}\le c\bigg\}\\
		&\quad+ \mathbb{P}\bigg\{\sup _{t_{n,i} \le t \le t_{n,i}+d_{n,i}}\frac{\Pi_{i_2}(t)}{a(t)} \ge 1,\ t_{n,i}<\infty,\ d_{n,i}\le c\bigg\},\\
		&\overset\Delta=\Theta_{i_1}(t)+\Theta_{i_2}(t).\\
	\end{aligned}
\end{equation}
Next, we will estimate $\Theta_{i_1}(t)$ and $\Theta_{i_2}(t)$ respectively.

To this end, we first define  $\breve{x}_{ji}(t)=\int_{t_{k,i}}^{t}\tilde{x}_{ji}(t)a(t)dt$, where $\tilde{x}_{ji}(t)=\|\tilde{x}_{j}(t)-\tilde{x}_{i}(t)\|$. Then, it is easy to get
\begin{equation*}
	\begin{aligned}
		\Theta_{i_1}(t)
		=&\mathbb{P}\bigg\{\sup _{t_{n,i} \le t \le t_{n,i}+d_{n,i}}4N^3\frac{\sum_{j \in \mathcal{N}_{i}} a_{i j}^4\breve{x}_{ji}^4(t)}{a^2(t)} \ge 1,\\
		&\qquad\qquad\qquad\qquad\qquad \quad t_{n,i}<\infty,\ d_{n,i}\le c\bigg\}\\
	\end{aligned}
\end{equation*}
and
\begin{equation}
	\begin{aligned}\label{brevexji}
		&\quad \sup _{t_{n,i} \le t \le t_{n,i}+d_{n,i}}\frac{\sum_{j \in \mathcal{N}_{i}} a_{i j}^4\breve{x}_{ji}^4(t)}{a^2(t)}\\
		& \le \frac{\sum_{j \in \mathcal{N}_{i}} a_{i j}^4 \big(\int 	_{t_{n,i}}^{t_{n,i}+d_{n,i}}\tilde{x}_{ji}(t)a(t)  d t\big)^4}{a^2(t_{n,i}+d_{n,i})} \\
		& \le \sum_{j \in \mathcal{N}_{i}} a_{i j}^4 \frac{ \big(\int _{t_{n,i}}^{t_{n,i}+d_{n,i}}\tilde{x}_{ji}^2(t)dt\big)^2\big( \int _{t_{n,i}}^{t_{n,i}+d_{n,i}}a^2(t)  d t\big)^2}{a^2(t_{n,i}+d_{n,i})} \\
		& \le\sum_{j \in \mathcal{N}_{i}} a_{i j}^4 \frac{d_{n,i} \int _{t_{n,i}}^{t_{n,i}+d_{n,i}}\tilde{x}_{ji}^4(t)dt\big( \int _{t_{n,i}}^{t_{n,i}+d_{n,i}}a^2(t)  d t\big)^2}{a^2(t_{n,i}+d_{n,i})}, \\
	\end{aligned}
\end{equation}
where the first inequality follows from the monotonicity of $\frac{\sum_{j \in \mathcal{N}_{i}} a_{i j}^4\breve{x}_{ji}^4(t)}{a^2(t)}$, the second and third inequality take advantage of the Cauchy-Schwarz inequality.

Note that $d_{n,i} \le \frac{t_{n,i}}{n} < \frac{t_{n,i}+1}{n}$, then we have
\begin{equation}\label{ine_3}
	\begin{aligned}
		&\quad\frac{\Big( \int _{t_{n,i}}^{t_{n,i}+d_{n,i}}a^2(t)  d t\Big)^2}{a^2(t_{n,i}+d_{n,i})}\\
		& \le \frac{(1+t_{n,i})^{-2\gamma}d_{n}^2}{(1+t_{n,i}+d_{n,i})^{-2\gamma}}\\
		& = \Big(\frac{1+t_{n,i}+d_{n,i}}{1+t_{n,i}}\Big)^{2\gamma}\Big(\frac{d_{n,i}}{1+t_{n,i}}\Big)^{2\gamma}d_{n,i}^{2-2\gamma}\\
		&\le 2^{2\gamma}d_{n,i}^{2-2\gamma}\frac{1}{n^{2\gamma}}.
	\end{aligned}
\end{equation}

Substituting \eqref{ine_3} into \eqref{brevexji}, we can get that

\begin{equation*}
	\begin{aligned}
		&\quad\sup _{t_{n,i} \le t \le t_{n,i}+d_{n,i}}\frac{\sum_{j \in \mathcal{N}_{i}} a_{i j}^4\breve{x}_{ji}^4(t)}{a^2(t)}\\
		& \le 4^{\gamma} d_{n,i}^{3-2\gamma} \frac{1}{n^{2\gamma}}\sum_{j \in \mathcal{N}_{i}} a_{i j}^4 \int _{t_{n,i}}^{t_{n,i}+d_{n,i}}\tilde{x}_{ji}^4(t)dt.\\
	\end{aligned}	
\end{equation*}
Therefore,
\begin{equation*}
	\begin{aligned}
		&\Theta_{i_1}(t)\\
		\le & \mathbb{P}\bigg\{4^{\gamma+1}N^3 d_{n,i}^{3-2\gamma} \frac{1}{n^{2\gamma}}\sum_{j \in \mathcal{N}_{i}} a_{i j}^4 \int _{t_{n,i}}^{t_{n,i}+d_{n,i}}\tilde{x}_{ji}^4(t)dt \ge 1,\\
		&\qquad\qquad \qquad \qquad \qquad \qquad \qquad  t_{n,i}<\infty,\ d_{n,i}\le c\bigg\}\\
		\le & \mathbb{P}\bigg\{4^{\gamma+1}N^3 c^{3-2\gamma} \frac{1}{n^{2\gamma}}\sum_{j \in \mathcal{N}_{i}} a_{i j}^4 \int _{t_{n,i}}^{t_{n,i}+d_{n,i}}\tilde{x}_{ji}^4(t)dt \ge 1,\\ &\qquad\qquad \qquad \qquad \qquad \qquad \qquad   t_{n,i}<\infty,\ d_{n,i}\le c\bigg\}\\
		\le & 4^{\gamma+1}N^3 c^{3-2\gamma} \frac{1}{n^{2\gamma}}\sum_{j \in \mathcal{N}_{i}} a_{i j}^4 \mathbb{E}\bigg(\int _{t_{n,i}}^{t_{n,i}+d_{n,i}}\tilde{x}_{ji}^4(t)dt\bigg)\\
		\le & 4^{\gamma+1}N^3 c^{3-2\gamma} \frac{1}{n^{2\gamma}}\sum_{j \in \mathcal{N}_{i}} a_{i j}^4 \mathbb{E}\bigg(\int _{0}^{\infty}\tilde{x}_{ji}^4(t)dt\bigg)\\
		= & 4^{\gamma+1}N^3 c^{3-2\gamma} \frac{1}{n^{2\gamma}}\sum_{j \in \mathcal{N}_{i}} a_{i j}^4 \int _{0}^{\infty}\mathbb{E}(\tilde{x}_{ji}^4(t))dt.
	\end{aligned}
\end{equation*}

From (\ref{rate1}) we know
\begin{equation}\label{rate}
	\mathbb{E}V(t) \le C_{2,p} t^{-\frac{p\gamma}{2}},
\end{equation}
which means there exists a constant $\tilde{C}>0$, such that for $\forall 1\le i,j \le N$ and $\forall t>0$, the following inequality holds:
\begin{equation}\label{expb}
	\mathbb{E}\left\|\tilde{x}_{j}(t)-\tilde{x}_{i}(t)\right \|^4 \le \tilde{C} (t+1)^{-2\gamma}.
\end{equation}
It is easy to figure out
\begin{equation}
	\int _{0}^{\infty}\mathbb{E}\left\|\tilde{x}_{j}(t)-\tilde{x}_{i}(t)\right \|^4 \le \frac{\tilde{C}}{2\gamma-1}.
\end{equation}
Therefore,
\begin{equation}\label{esti 1}
	\Theta_{i_1}(t)	\le \bigg(4^{\gamma+1}N^3 c^{3-2\gamma} \frac{\tilde{C}}{2\gamma-1}\sum_{j \in \mathcal{N}_{i}} a_{i j}^4\bigg) \frac{1}{n^{2\gamma}}.
\end{equation}
\par Secondly, for $\Theta_{i_2}(t)$, we first prove that $\exists$ a constant $ M_{1}>0$, such that
\begin{equation}\label{P_1}
	\begin{aligned}
		&\mathbb{P}\Bigg\{\sup _{t_{n,i} \le t \le t_{n,i}+d_{n,i}}\frac{\Pi_{i_2}(t)}{a(t)} \ge 1|\mathcal{F}_{t_{n,i}}\Bigg\}\\
		\le& 4M_{1} N^3\frac{\sum_{j \in \mathcal{N}_{i}} a_{i j}^4 \sigma_{j i}^4\big(\int _{t_{n,i}}^{t_{n,i}+d_{n,i}}a^2(t)dt\big)^2}{a^2(t_{n,i}+d_{n,i})}.
	\end{aligned}
\end{equation}
\par By Chebyshev inequality for conditional expectation and monotonicity of $a(t)$, we have
\begin{equation}\label{P_2}
	\begin{aligned}
		&\mathbb{P}\Bigg\{\sup _{t_{n,i} \le t \le t_{n,i}+d_{n,i}}\frac{\Pi_{i_2}(t)}{a(t)} \ge 1|\mathcal{F}_{t_{n,i}}\Bigg\}\\
		\le & \mathbb{E}\Bigg(\sup _{t_{n,i} \le t \le t_{n,i}+d_{n,i}}\frac{\Pi_{i_2}^2(t)}{a^2(t)} |\mathcal{F}_{t_{n,i}}\Bigg) \\
		\le &\frac{\mathbb{E}\big(\sup _{t_{n,i} \le t \le t_{n,i}+d_{n,i}}\Pi_{i_2}^2(t)|\mathcal{F}_{t_{n,i}}\big)}{a^2(t_{n,i}+d_{n,i})}. \\
	\end{aligned}
\end{equation}
\par Define 
\begin{equation}
	X_{n,i}(t)\overset\Delta=\sum_{j \in \mathcal{N}_{i}} a_{i j} \sigma_{j i}\int _{0}^{t}a(r) 1_{r\ge t_{n,i}}dw_{ji}(r).
\end{equation}
It is a continuous martingale with $X_{n,i}(0)=0$.
Therefore, by conditional Burkholder-Davis-Gundy inequality (see Theorem 2.3.8 in \cite{Zhou21}), $\exists$ a constant $ M_{1}>0$, such that
\begin{equation}
	\begin{aligned}
		& \mathbb{E}\bigg(\sup _{t_{n,i} \le t \le t_{n,i}+d_{n,i}}X_{n,i}^4(t)|\mathcal{F}_{t_{n,i}}\bigg)\\
		\le & M_{1} \mathbb{E}(\left \langle X_{n,i},X_{n,i} \right \rangle_{(t_{n,i}+d_{n,i})} ^2|\mathcal{F}_{t_{n,i}}) \ \text{a.s.},
	\end{aligned}
\end{equation}
where $\left \langle X_{n,i},X_{n,i} \right \rangle$ is the quadratic variation of $X_{n,i}$. Thus,
\begin{equation}\label{EPi_{i_2}}
	\begin{aligned}
		&\mathbb{E}\Bigg(\sup _{t_{n,i} \le t \le t_{n,i}+d_{n,i}}\Pi_{i_2}^2(t)|\mathcal{F}_{t_{n,i}}\Bigg)\\
		=&4	\mathbb{E}\bigg(\sup _{t_{n,i} \le t \le t_{n,i}+d_{n,i}}X_{n,i}^4(t)|\mathcal{F}_{t_{n,i}}\bigg) \\
		\le & 4M_{1}  \mathbb{E}(\left \langle X_{n,i},X_{n,i} \right \rangle_{(t_{n,i}+d_{n,i})} ^2|\mathcal{F}_{t_{n,i}}) \\
		\le & 4M_{1}N^3\sum_{j \in \mathcal{N}_{i}} a_{i j}^4 \sigma_{j i}^4(\int _{t_{n,i}}^{t_{n,i}+d_{n,i}}a^2(t)dt)^2 \ \text{a.s.}.
	\end{aligned}
\end{equation}
Substituting \eqref{EPi_{i_2}} into \eqref{P_2}, it is easy to obtain \eqref{P_1}.

Then, by the property of conditional expectation, we have
\begin{equation}
	\begin{aligned}
		\Theta_{i_2}(t)= & \mathbb{E}\Bigg(\mathbb{P}\Bigg\{\sup _{t_{n,i} \le t \le t_{n,i}+d_{n,i}}\frac{\Pi_{i_2}(t)}{a(t)} \ge 1,\\
		&\qquad \qquad \qquad \qquad \  t_{n,i}<\infty,\ d_{n,i}\le c|\mathcal{F}_{t_{n,i}}\Bigg\}\Bigg)\\
		= & \mathbb{E}\Bigg(\mathbb{P}\Bigg\{\sup _{t_{n,i} \le t \le t_{n,i}+d_{n,i}}\frac{\Pi_{i_2}(t)}{a(t)} \ge 1|\mathcal{F}_{t_{n,i}}\Bigg\}\\
		&\qquad \qquad \qquad \qquad \qquad \qquad \quad 1_{t_{n,i}<\infty,\ d_{n,i}\le c}\Bigg)\\
		\le & \mathbb{E}\Bigg(4M_{1} N^3\frac{\sum_{j \in \mathcal{N}_{i}} a_{i j}^4 \sigma_{j i}^4\big(\int _{t_{n,i}}^{t_{n,i}+d_{n,i}}a^2(t)dt\big)^2}{a^2(t_{n,i}+d_{n,i})}\\
		& \qquad \qquad \qquad \qquad \qquad \qquad \quad 1_{t_{n,i}<\infty,\ d_{n,i}\le c}\Bigg)\\
		\le & \mathbb{E}\Bigg(4M_{1}N^3\frac{\sum_{j \in \mathcal{N}_{i}} a_{i j}^4 \sigma_{j i}^4\big(d_{n,i}(1+t_{n,i})^{-2\gamma } \big)^2}{(1+t_{n,i}+d_{n,i})^{-2\gamma}}\\
		& \qquad \qquad \qquad \qquad \qquad \qquad \quad 1_{t_{n,i}<\infty,\ d_{n,i}\le c}\Bigg)\\
		= & \mathbb{E}\Bigg(4M_{1}N^3\sum_{j \in \mathcal{N}_{i}} a_{i j}^4 \sigma_{j i}^4 d_{n,i}^{2-2\gamma}\bigg(\frac{d_{n,i}}{1+t_{n,i}}\bigg)^{2\gamma}\\
		&\qquad \qquad \quad \bigg(\frac{1+t_{n,i}+d_{n,i}}{1+t_{n,i}}\bigg)^{2\gamma} 1_{t_{n,i}<\infty,d_{n,i}\le c}\Bigg)\\
		\le & \bigg(4^{\gamma+1 }M_{1}N^3 c^{2-2\gamma }\sum_{j \in \mathcal{N}_{i}} a_{i j}^4 \sigma_{j i}^4\bigg)\frac{1}{n^{2\gamma}},
	\end{aligned}
\end{equation}
where the first inequality follows from (\ref{P_1}), the second inequality follows from $a(t)=(1+t)^{-\gamma}$, the third inequality makes use of $d_{n,i} \le \frac{t_{n,i}}{n} < \frac{t_{n,i}+1}{n}$ like before.

Therefore,
\begin{equation} \label{esti 2}
	\Theta_{i_2}(t)	\le \frac{4^{\gamma+1 }M_{1}N^3 c^{2-2\gamma}}{n^{2\gamma}}\sum_{j \in \mathcal{N}_{i}} a_{i j}^4 \sigma_{j i}^4.
\end{equation}

Combining (\ref{esti 1}) and (\ref{esti 2}), we have
\begin{align*}
	&\sum_{n=1}^{\infty } \mathbb{P}\{A_{n,i}\} \\
	\le & \sum_{n=1}^{\infty}\big(\Theta_{i_1}(t)+\Theta_{i_2}(t)\big)\\
	\le & \bigg(4^{\gamma+1}N^3 c^{3-2\gamma} \frac{\tilde{C}}{2\gamma-1}\sum_{j \in \mathcal{N}_{i}} a_{i j}^4\\
	&+4^{\gamma+1 }M_{1}N^3 c^{2-2\gamma }\sum_{j \in \mathcal{N}_{i}} a_{i j}^4 \sigma_{j i}^4\bigg)\sum_{n=1}^{\infty }\frac{1}{n^{2\gamma}}\\
	< & \infty, \\
\end{align*}
i.e.,
\begin{equation}
	\sum_{n=1}^{\infty } \mathbb{P}\{A_{n,i}\} < \infty.
\end{equation}
\par According to Borel-Cantelli lemma,
\begin{equation}
	\mathbb{P}\{A_{n,i}\ \text{i.o.}\}=0.
\end{equation}
So
\begin{equation}
	\begin{aligned}
		&\mathbb{P}\{\inf_{k}( t_{k+1,i}-t_{k,i})=0\}\\
		=&\mathbb{P}\{\exists\ \text{a decreasing subsequence}\ \triangle t_{k_{l},i}\ \text{converges to 0} \}\\
		\le&\mathbb{P}\{A_{n,i}\ \text{i.o.}\}.
	\end{aligned}
\end{equation}
Therefore, the Zeno behavior can be excluded a.s..  $\hfill\blacksquare$

The above theorem proves that there is a strict positive lower bound between two adjacent execution times of each agent $i$, thus excluding Zeno behavior. To better clarify this conclusion, we give the following estimates.
\begin{corollary}
	For any constant $\iota>0$, the inequality
	\begin{equation}
		t_{n+1,i}-t_{n,i}> \iota
	\end{equation}
	holds with probability at least $1-\Big(\frac{4N^3\overline{C}^4\tilde{C}\sum_{j \in \mathcal{N}_{i}}a_{ij}^4}{\underline{C}^2(2\gamma-1)}\iota+\frac{4M_1N^3\overline{C}^4\sum_{j \in \mathcal{N}_{i}}a_{ij}^4\sigma_{ji}^4}{\underline{C}^2}\Big)\iota^2(1+\iota)^{2\gamma}$.
	
\end{corollary}
{\bf Proof}.

From (\ref{trigger}) we have
\begin{align*}
	&\quad \mathbb{P}\{t_{n+1,i}-t_{n,i}\le \iota\} \\
	&= \mathbb{P}\bigg\{\sup _{t_{n,i} \le t \le t_{n,i}+\iota}\frac{e^{T}_{i}(t)e_{i}(t)}{a(t)} \ge 1,\ t_{n,i}< \infty \bigg\}\\
	&\le  \mathbb{P}\bigg\{\sup _{t_{n,i} \le t \le t_{n,i}+\iota}\frac{\Pi_{i_1}(t)+\Pi_{i_2}(t)}{a(t)} \ge 1, t_{n,i}<\infty \bigg\} \\
	& \le \mathbb{P}\bigg\{\sup _{t_{n,i} \le t \le t_{n,i}+\iota}\frac{\Pi_{i_1}(t)}{a(t)} \ge 1,\ t_{n,i}<\infty\bigg\}\\
	&\quad+ \mathbb{P}\bigg\{\sup _{t_{n,i} \le t \le t_{n,i}+\iota}\frac{\Pi_{i_2}(t)}{a(t)} \ge 1,\ t_{n,i}<\infty\bigg\}\\
	&\overset\Delta=\Theta_{i_1,\iota}(t)+\Theta_{i_2,\iota}(t).\\
\end{align*}

And note that
\begin{equation*}
	\begin{aligned}
		&\frac{	\Big( \int _{t_{n,i}}^{t_{n,i}+\iota}a^2(t)  d t\Big)^2}{a^2(t_{n,i}+\iota)}\\
		\le&\frac{\overline{C}^4(1+t_{n,i})^{-4\gamma}\iota^2}{\underline{C}^2(1+t_{n,i}+\iota)^{-2\gamma}}\\
		=&\frac{\overline{C}^4}{\underline{C}^2}\Big(\frac{1+t_{n,i}+\iota}{1+t_{n,i}}\Big)^{2\gamma}\Big(\frac{\iota}{1+t_{n,i}}\Big)^{2\gamma}\iota^{2-2\gamma}\\
		\le& \frac{\overline{C}^4}{\underline{C}^2}\iota^2(1+\iota)^{2\gamma}
	\end{aligned}
\end{equation*}

Then, by proving method similar to (\ref{esti 1}) and (\ref{esti 2}), we can obtain the following two inequalities:
\begin{equation}\label{cor ineq1}
	\Theta_{i_1,\iota}(t)\le \frac{4N^3\overline{C}^4\tilde{C}\sum_{j \in \mathcal{N}_{i}}a_{ij}^4}{\underline{C}^2(2\gamma-1)}\iota^3(1+\iota)^{2\gamma},
\end{equation}
\begin{equation}\label{cor ineq2}
	\Theta_{i_2,\iota}(t)\le \frac{4M_1N^3\overline{C}^4\sum_{j \in \mathcal{N}_{i}}a_{ij}^4\sigma_{ji}^4}{\underline{C}^2}\iota^2(1+\iota)^{2\gamma}.
\end{equation}

Therefore,
\begin{align*}
	&\quad \mathbb{P}\{t_{n+1,i}-t_{n,i}\le \iota\} \\	& \le  \Theta_{i_1,\iota}(t)+\Theta_{i_2,\iota}(t)\\
	&\le \frac{4N^3\overline{C}^4\tilde{C}\sum_{j \in \mathcal{N}_{i}}a_{ij}^4}{\underline{C}^2(2\gamma-1)}\iota^3(1+\iota)^{2\gamma}\\&\quad+\frac{4M_1N^3\overline{C}^4\sum_{j \in \mathcal{N}_{i}}a_{ij}^4\sigma_{ji}^4}{\underline{C}^2}\iota^2(1+\iota)^{2\gamma}\\
\end{align*}
which means the inequality 
\begin{equation}
	t_{n+1,i}-t_{n,i}> \iota
\end{equation}
holds with probability at least $1-(\frac{4N^3\overline{C}^4\tilde{C}\sum_{j \in \mathcal{N}_{i}}a_{ij}^4}{\underline{C}^2(2\gamma-1)}\iota+\frac{4M_1N^3\overline{C}^4\sum_{j \in \mathcal{N}_{i}}a_{ij}^4\sigma_{ji}^4}{\underline{C}^2})\iota^2(1+\iota)^{2\gamma}$. $\hfill\blacksquare$

\section{Simulation} \label{Sim}
\begin{figure}[htbp]
	\centering
	\includegraphics[width=0.7\columnwidth]{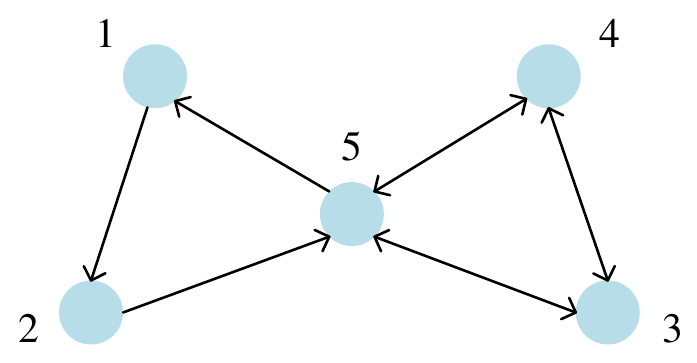}
	\caption{The communication topology.}
	\label{to}
\end{figure}

In this section, a numerical simulation is presented to illustrate the effectiveness of the proposed distributed event-triggered privacy preserving weighted consensus scheme. The communication topology of the MAS is shown in Fig. \ref{to}. The initial conditions of the agents are $x_1(0)=5$, $x_2(0)=3$, $x_3(0)=1$, $x_4(0)=2$, and $x_5(0)=4$. The parameters are choosen as $\gamma=0.6$, $k=0.01$ and $h=0.3$.

\begin{figure}[htbp]
	\centering
	\includegraphics[width=0.66\columnwidth]{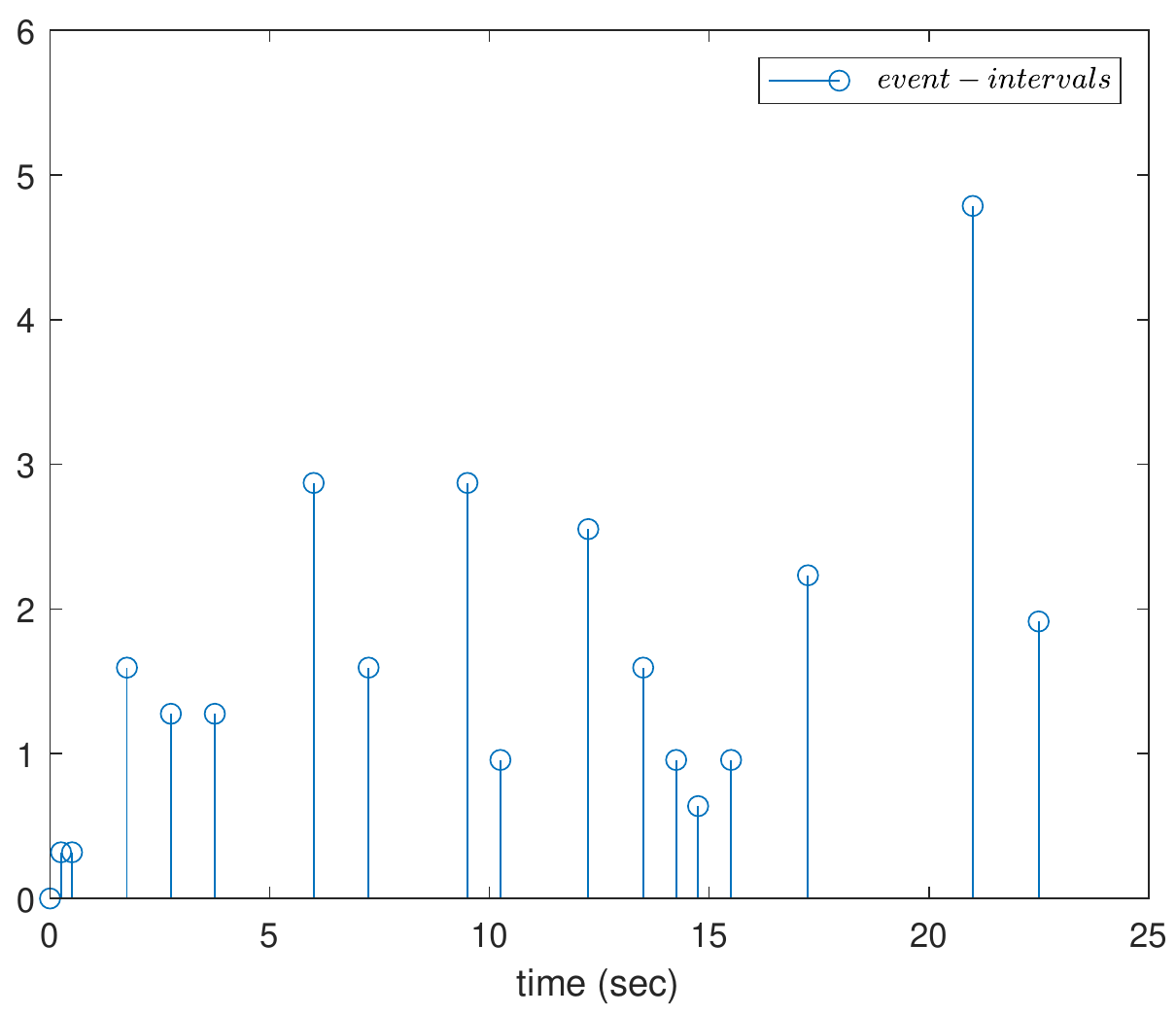}
	\caption{Event-triggering intervals of agent 1.}
	\label{ev}
\end{figure}

We first verify the effectiveness of the proposed event triggering mechanism. Take agent 1 as an example, and its event triggering interval is shown in Fig. \ref{ev}, where the abscissa and ordinate represent the time instant and the time interval between two adjacent events, respectively. Fig. \ref{ev} shows that the time interval between two adjacent events has a consistent lower bound, which indicates that Zeno behavior can be ruled out.

\begin{figure}[htbp]
	\centering
	\includegraphics[width=0.7\columnwidth]{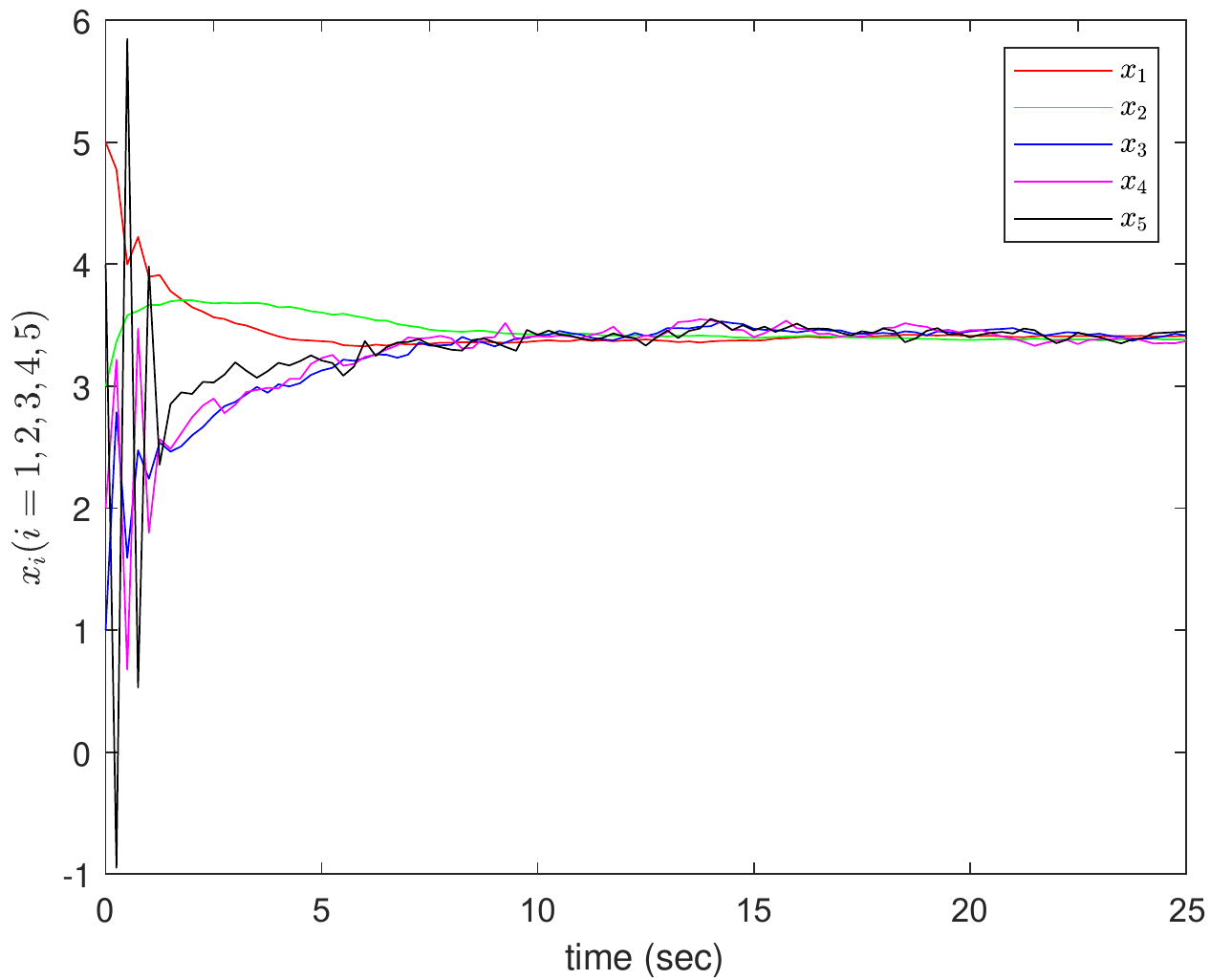}
	\caption{The state trajectory of the agents.}
	\label{st}
\end{figure}

\begin{figure}[htbp]
	\centering
	\includegraphics[width=0.7\columnwidth]{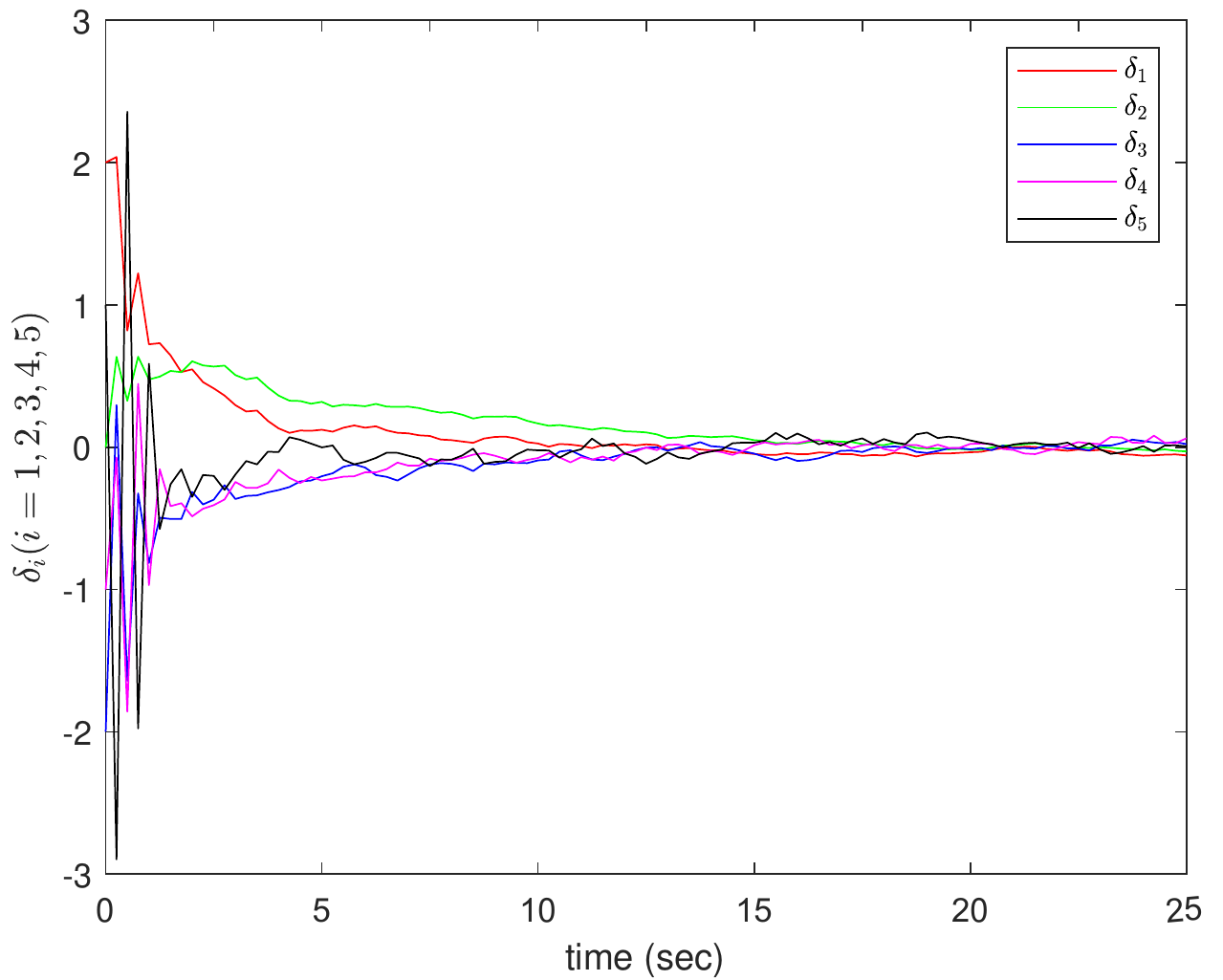}
	\caption{The consensus error of the agents.}
	\label{et}
\end{figure}

\begin{figure}[htbp]
	\centering
	\includegraphics[width=0.7\columnwidth]{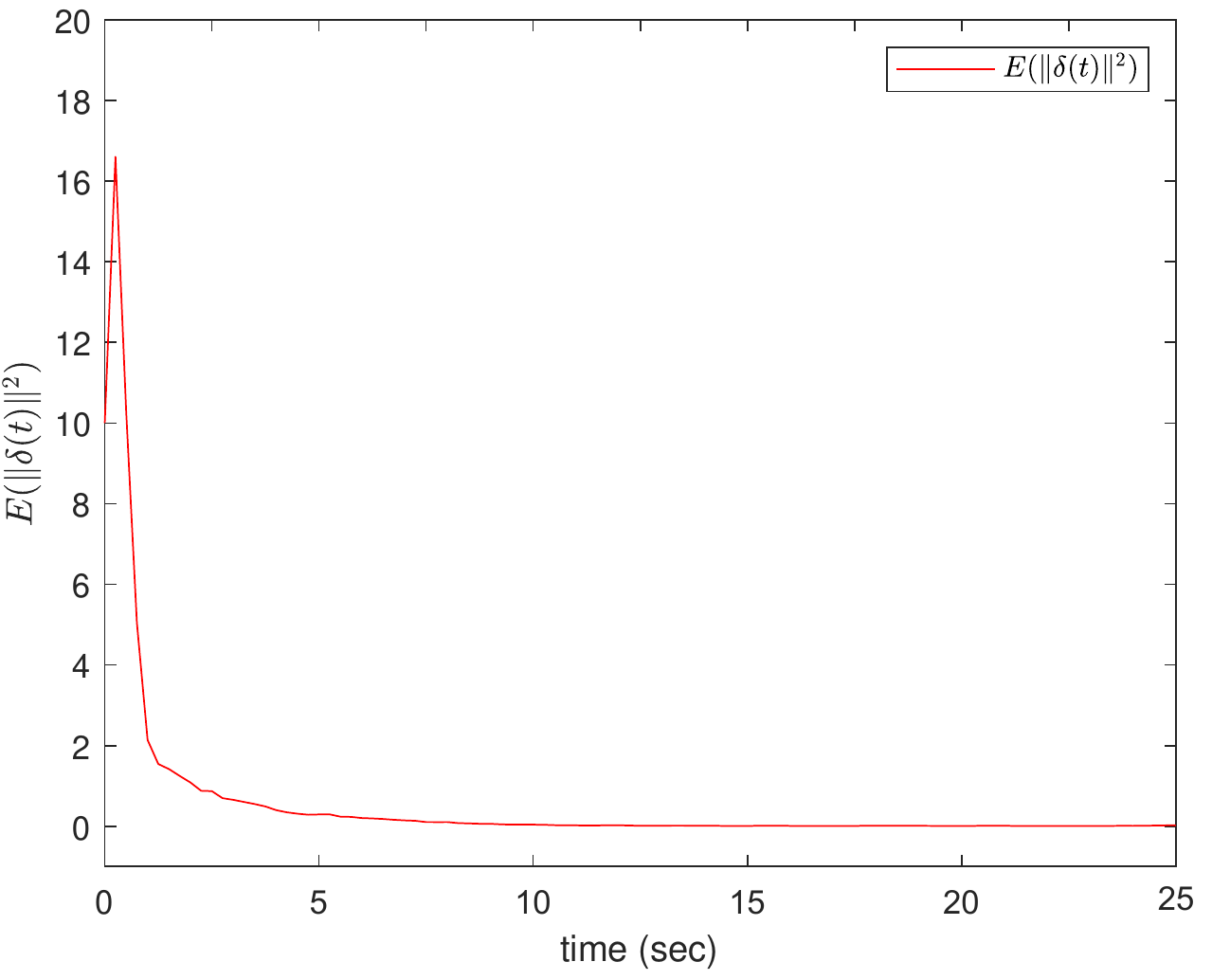}
	\caption{The mean square of the consensus error.}
	\label{E}
\end{figure}

Fig. \ref{st} and Fig. \ref{et} show the state trajectory and consensus error of each agent, respectively. As can be seen from these figures, since privacy protection is taken into account in the consensus control algorithm, each agent cannot converge to the  weighted average of the initial states deterministically. However, it can be seen from Fig. \ref{E} that the mean-square convergence of consensus error can be achieved. These results show that our proposed privacy preserving consistency control algorithm is effective.

\section{Conclusion} \label{Con}
In this paper, we proposed a distributed event-triggered privacy preserving weighted consensus control scheme for MASs. In order to reduce unnecessary communication loss, we have designed a novel event triggering mechanism for each agent to determine the information transmission instants. Then, we have developed a privacy preserving scheme, which protects the state information of each agent from disclosure by adding edge-based mutually independent standard white noises to the communication process with neighbors. Further, we proposed a stochastic approximation type protocol for each agent to attenuates the effect of noises on consensus control. Using the tools of stochastic analysis and algebraic theory, the asymptotic property and convergence accuracy of consensus error is analyzed. Theoretical analysis and numerical simulation verify that the proposed method can successfully make privacy preserving weighted consensus without incurring Zeno behavior.

\section*{Appendix}
\subsection*{Proof of Lemma \ref{a1}}
According to Assumption \ref{assumption2}, there is $t_0>0$ and $\epsilon>0$ such that $\forall t\ge t_0$, $a(t)\ge\epsilon t^{-\gamma}$. Thus, it yields that
\begin{equation*}
	e^{-\mu\int_0^ta(s)ds}\le e^{-\mu\int_0^{t_{0}}a(s)ds}e^{\frac{\mu\epsilon}{1-\gamma}t_0^{1-\gamma}}e^{-\frac{\mu\epsilon}{1-\gamma}t^{1-\gamma}}.
\end{equation*}

Let $\nu=t^{1-\gamma}$. Since $\gamma\in(0.5,1)$, then as $t\to\infty$, $\nu\to\infty$. And because of $\frac{\gamma}{1-\gamma}>0$ and $-\frac{\mu\epsilon}{1-\gamma}<0$, we can get that
\begin{equation*}
	\lim_{t\to\infty}t^{\frac{p\gamma}{2}}e^{-\frac{\mu\epsilon}{1-\gamma}t^{1-\gamma}}=\lim_{\nu\to\infty}\nu^{\frac{2\gamma}{p(1-\gamma)}}e^{-\frac{\mu\epsilon}{1-\gamma}\nu}=0.
\end{equation*}
Thus,
\begin{equation*}
	\lim_{t\to\infty}t^{\frac{p\gamma}{2}}e^{-\mu\int_0^ta(s)ds}=0,
\end{equation*}
i.e.,
\begin{equation*}
	e^{-\mu\int_0^ta(s)ds}=o(t^{-\frac{p\gamma}{2}})
\end{equation*}

Further, according to Assumption \ref{assumption2} and \eqref{lemma2_1}, using L'Hôpital's rule, it follows that
\begin{equation*}
	\begin{aligned}
		&\lim_{t\to\infty}t^{\frac{p\gamma}{2}}\int_0^t a^2(s)e^{-\mu\int_s^ta(r)dr}ds\\
		=&\lim_{t\to\infty}\frac{d(\int_0^t a^2(s)e^{\mu\int_0^sa(r)dr}ds)/dt}{d(t^{-\frac{p\gamma}{2}} e^{\mu\int_0^ta(r)dr})/dt}\\
		=&\frac{(\lim_{t\to\infty}t^{\gamma}a(t))^{\frac{p}{2}}}{\mu}.
	\end{aligned}
\end{equation*}
This completes the proof. $\hfill\blacksquare$


\begin{thebibliography}{unsrt}
	\bibitem{Tas22}
T. K. Tasooji and H. J. Marquez, Event-triggered consensus control for multi-robot systems with cooperative localization, IEEE Transactions on Industrial Electronics, 2022.

\bibitem{Xu20}
Y. Xu, T. Li, and S. Tong, Event-triggered adaptive fuzzy bipartite consensus control of multiple autonomous underwater vehicles, IET Control Theory \& Applications, 14(20), 3632--3642, 2020.

\bibitem{Zhao18}
C. Zhao, J. Chen, J. He, and P. Cheng, Privacy-preserving consensus-based energy management in smart grids, IEEE Transactions on Signal Processing, 66(23), 6162--6176, 2018.

\bibitem{Dong15}
X. Dong, B. Yu, Z. Shi, and Y. Zhong, Time-varying formation control for unmanned aerial vehicles: Theories and applications, IEEE Transactions on Control Systems Technology, 23(1), 340--348, 2015.

\bibitem{Qin17}
J. Qin, Q. Ma, Y. Shi, and L. Wang, Recent advances in consensus of multi-agent systems: A brief survey, IEEE Transactions on Industrial Electronics, 64(6), 4972--4983, 2017.

\bibitem{Ma17}
L. Ma, Z. Wang, Q.-L. Han, and Y. Liu, Consensus control of stochastic multi-agent systems: A survey, Science China. Information sciences, 60(12), 1-15, 2017.

\bibitem{Ding18}
L. Ding, Q.-L. Han, X. Ge, and X.-M. Zhang, An overview of recent avances in event-triggered consensus of multiagent systems, IEEE Transactions on Aerospace and Electronic Systems, 48(4), 1110--1123, 2018.

\bibitem{Ruan19}
M. Ruan, H.	Gao, and Y.	Wang, Secure and privacy-preserving consensus, IEEE Transactions on Automatic Control, 64, 4035--4049, 2019.

\bibitem{Had20}
C. N. Hadjicostis and A. D. Dominguez-Garcia, Privacy-Preserving Distributed Averaging via Homomorphically Encrypted Ratio Consensus, IEEE Transactions on Automatic Control, 65(9), 3887--3894, 2020.

\bibitem{Wang19}
Y. Wang, Privacy-preserving average consensus via state decomposition, IEEE Transactions on Automatic Control, 64(11), 4711--4716, 2019.

\bibitem{Zhang22}
K. Zhang, Z. Li, Y. Wang, A. Louati, and J. Chen, Privacy-preserving dynamic average consensus via state decomposition: Case study on multi-robot formation control, Automatica, 139, 110182, 2022.

\bibitem{Hu22}
J. Hu, Q. Sun, R. Wang, B. Wang, M. Zhai, and H. Zhang, Privacy-preserving sliding mode control for voltage restoration of AC microgrids based on output mask approach, IEEE Transactions on Industrial Informatics, 18(10), 6818--6827, 2022.

\bibitem{Mo17}
Y. Mo and R. M. Murray, Privacy preserving average consensus, IEEE Transactions on Automatic Control, 62(2), 753--765, 2017.

\bibitem{Xiong22}
Y. Xiong and Z.	Li, Privacy-preserved average consensus algorithms with edge-based additive perturbations, Automatica, 140, 110223, 2022.

\bibitem{Gao19}
L. Gao, S. Deng, and W. Ren, Differentially private consensus with an event-triggered mechanism, IEEE Transactions on Control of Network Systems, 6(1), 60--71, 2019.

\bibitem{Wang21}
A. Wang, H.	He, and X. Liao, Event-triggered privacy-preserving average consensus for multiagent networks with time delay: An output mask approach, IEEE Transactions on Systems, Man, and Cybernetics. Systems, 51(7), 4520--4531, 2021.

\bibitem{Yang22}
Z. Yang, L.	Yu, Y. Liu, N. D. Alotaibi, and F. E. Alsaadi, Event-triggered privacy-preserving bipartite consensus for multi-agent systems based on encryption, Neurocomputing, 503, 162--172, 2022.

\bibitem{Nozari17}
E. Nozari, P. Tallapragada, and J. Cort\'{e}s, Differentially private average consensus: Obstructions, trade-offs, and optimal algorithm design, Automatica, 81, 221--231, 2017.

\bibitem{Fiore19}
D. Fiore and G. Russo, Resilient consensus for multi-agent systems subject to differential privacy requirements, Automatica, 106, 18--26, 2019.

\bibitem{Wang19}
X. Wang, J. He, P. Cheng, J. Chen, Differentially private maximum consensus: Design, analysis and impossibility result, IEEE Transactions on Network Science and Engineering, 6(4), 928--939, 2019.

\bibitem{He20}
J. He, L. Cai, and X. Guan, Differential private noise adding mechanism and its application on consensus algorithm, IEEE Transactions on Signal Processing, 68, 4069-4082, 2020.

\bibitem{Dong20}
T. Dong, X. Bu, and W. Hu, Distributed differentially private average consensus for multi-agent networks by additive functional Laplace noise, Journal of the Franklin Institute, 357(6), 3565--3584, 2020.

\bibitem{Yang16}
D. Yang, W. Ren, X. Liu, and W.	Chen, Decentralized event-triggered consensus for linear multi-agent systems under general directed graphs, Automatica, 69, 242--249, 2016.

\bibitem{Zhou21}
M.Zhou, Contributions to the numerics of quadratic backward SDEs and probability on turbulence,PhD Thesis,St Hugh's College,Univ. Oxford,Oxford,UK, 2021.
\end{thebibliography}
\end{document}